\documentclass[preprint,preprintnumbers,amsmath,amssymb]
{revtex4}

\usepackage{graphicx}
\usepackage{dcolumn}
\usepackage{bm}

\begin{document}

\title{Hydrodynamic Implosion Simulation Including
Relativistic Effects on Petawatt-Class Pulse Heating}

\author{Mitsuru~Honda}
\affiliation{Plasma Astrophysics Laboratory, Institute for Global Science,
Mie 519-5203, Japan}

\begin{abstract}
A spherically symmetric fluid code that includes the
relativistic effects of hot electrons has been newly developed.
In the present simulations, I attempt to implode
a target shell of laser fusion using a nanosecond pulse;
then, an additional heating short pulse is injected
into the coronal plasma surrounding a highly compressed core.
I found that the relativistic effects of hot electrons on electron transport
are likely to inhibit the heat flows, and
to reduce thermonuclear fusion neutrons significantly.
This feature may be important for off-center fast ignition
and burn of fusion targets.\\


\end{abstract}

\maketitle

The relativistic effects of hot electrons in laboratory plasmas have
attracted much interests for the past decades \cite{fisch87},
particularly, relating to laser fusion fast ignition (LFFI)
whose scenario is achieved by standard high-compression,
irradiation of a hole-boring laser beam,
irradiation of an ignitor laser beam, and thermonuclear burn \cite{tabak94}.
When high-intensity short pulse laser light is
impinged on the coronal plasma surrounding a highly compressed core
in the 'hole-boring' and 'ignition' stages,
energetic particles are sprayed through various mechanisms.
In order to investigate the relativistic electron transport
carrying huge currents, substantial studies were performed on the
multidimensional relativistic particle-in-cell \cite{pukhov97} as well as
magneto-hydrodynamic \cite{mason98} simulations.

Moreover, the fast ignition conditions of the compressed targets
have been presented in some publications
\cite{caruso96, atzeni97, piriz98a, piriz98b}.
Indeed, it is important to survey the energy transport
and relaxation in dense plasmas over a large spatiotemporal system of
$R > 10^1-10^2~{\rm \mu m}$ and $t > 1~{\rm ps}$.
Concerning the extension of the analysis of ignition conditions,
we are now strongly encouraged to improve the
transport coefficients so as to compare them with those for
kinetic simulations \cite{honda00, sentoku02} and
experiments \cite{tanaka00, kodama01}.
That is why, I attempted to construct a computational fluid code
that includes relativistic effects of electron transport.
In this code, the transport properties of hot electrons
are fully consistent with the current-neutral electric fields
self-induced in plasmas,
as discussed in ref.~\cite{honda98a} in detail.

In this paper, I argue that
in the context of the LFFI, heat flux inhibition
owing to relativistic effects of electrons degrades the
transport efficiency of deflagration thermal waves, and
leads to some reductions in the neutron yields.
This is essentially because
the drift velocity carrying heat asymptotically approaches the speed of light.
Since the velocity moment of the heat flux is large,
the energy of electrons carrying dominant heat (EE-CDH) is much larger than the
thermal energy: $T_{\rm EE-CDH} \approx 7 T_{\rm th}$.
Thus, it is expected that the relativistic effects of electron heat transport
can be observed even in the energy domain of
$T_{\rm th} \sim 10^1-10^2~{\rm keV}$.
It is noteworthy that such situations can be easily established
by irradiating a moderate-intensity laser light of
$I_{L} \lambda_{L}^2 \geq 10^{16}~{\rm W/cm^2-\mu m^2}$ \cite{honda98b}.
As for astrophysical aspects, the issue presented here is
relevant to supernova outbursts \cite{longair94}.

The spherically symmetric fluid code has been developed on the
Lagrangian frame ${\rm d}m=4 \pi \rho r^2 {\rm d}r$ with the
specific volume of $V=\rho^{-1}$.
This solves a set of two-temperature fluid equations for
an ideal gas, including the relativistically corrected equations for
the internal energy of electrons. That is,

\begin{widetext}
\begin{equation}
{\frac {{\rm d}u_i}{{\rm d}t}}=
-4 \pi r^2 {\frac {\partial}{\partial m}}
\left( P_e + P_i + P_{ph} \right),
\label{eq:1}
\end{equation}
\end{widetext}

\begin{widetext}
\begin{equation}
{\frac {{\rm d}\left( c_{ve\_rel}T_e  \right)}{{\rm d}t}}
+P_e {\frac {{\rm d}V}{{\rm d}t}} =
{\frac {\partial}{\partial m}}
\left( 4 \pi r^2 \kappa_{e\_rel}
{\frac {\partial T_e}{\partial r}} \right)
-\omega_{ei\_rel} \left( T_e-T_i \right)
- S_{brems} + S_{e\_nucl} + S_{L},
\label{eq:2}
\end{equation}
\end{widetext}

\begin{widetext}
\begin{equation}
c_{vi}{\frac {{\rm d}T_i}{{\rm d}t}}
+P_i {\frac {{\rm d}V}{{\rm d}t}} =
{\frac {\partial}{\partial m}}
\left( 4 \pi r^2 \kappa_i
{\frac {\partial T_i}{\partial r}} \right)
-\omega_{ei\_rel} \left( T_i-T_e \right) + S_{i\_nucl}.
\label{eq:3}
\end{equation}
\end{widetext}

\noindent\\
Here, conventional abbreviations have been used.
In particular, $\kappa_{e\_rel}$, $\omega_{ei\_rel}$, and
$c_{ve\_rel}$ denote the relativistic Spitzer-H\"arm
heat conductivity \cite{honda98a}, energy relaxation coefficient
\cite{beliaev56}, and specific heat \cite{balescu75}, respectively.
These coefficients normalized by their nonrelativistic values
are given by

\begin{widetext}
\begin{equation}
{\tilde{\kappa}_{e\_rel}}(\alpha) \equiv
{\frac {\kappa_{e\_rel}(\alpha)}{\kappa_{e\_nonrel}(\alpha)}}=
{\frac {\left( 2 \pi \right)^{1/2}}{384}}
{\frac {\alpha^{7/2}}{K_2(\alpha)}}
\left[ {\frac {\Theta_1^2(\alpha)}{\Theta_2(\alpha)}}
+ \Theta_3(\alpha) \right],
\label{eq:4}
\end{equation}
\end{widetext}

\begin{widetext}
\begin{equation}
{\tilde{\omega}_{ei\_rel}(\alpha)} \equiv
{\frac {\omega_{ei\_rel}(\alpha)}{\omega_{ei\_nonrel}(\alpha)}}=
{\frac {\left( 2 \pi \right)^{1/2}}{2}}
{\frac {{\rm exp}(-\alpha)}{\alpha^{1/2}K_2(\alpha)}}
\left( 1 + {\frac{2}{\alpha}} + {\frac{2}{\alpha^2}} \right),
\label{eq:5}
\end{equation}
\end{widetext}

\begin{widetext}
\begin{equation}
{\tilde c_{ve\_rel}(\alpha)} \equiv
{\frac {c_{ve\_rel}(\alpha)}{c_{ve\_nonrel}}}=
{\frac {2}{3}} \left[
\alpha^2 + 5 \alpha {\frac {K_3(\alpha)}{K_2(\alpha)}}
- \alpha^2 {\frac {K_3^2(\alpha)}{K_2^2(\alpha)}} - 1 \right],
\label{eq:6}
\end{equation}
\end{widetext}

\noindent\\
where $K_{\nu}(\alpha)$ is the modified Bessel function of
index $\nu$ with its argument of $\alpha \equiv m_0 c^2/T_e$,
and the functions $\Theta_1(\alpha)$, $\Theta_2(\alpha)$, 
and $\Theta_3(\alpha)$ are defined by \cite{honda03}

\begin{subequations}
\label{eq:7}
\begin{equation}
\Theta_1(\alpha)=\left(
1 - {1\over \alpha} + {2\over \alpha^2}
+ {42\over \alpha^3} + {120\over \alpha^4}
+ {120\over \alpha^5} \right) {\rm exp}(-\alpha)
+ \alpha {\rm Ei}(-\alpha),\label{subeq:a}
\end{equation}

\begin{equation}
\Theta_2(\alpha)=\left(
1 - {1\over \alpha} + {2\over \alpha^2}
- {6\over \alpha^3} - {24\over \alpha^4} - {24\over \alpha^5}
\right) {\rm exp}(-\alpha)
+ \alpha {\rm Ei}(-\alpha),\label{subeq:b}
\end{equation}

\begin{equation}
\Theta_3(\alpha)=\left(
{48\over \alpha^2} + {288\over \alpha^3}
+ {720\over \alpha^4} + {720\over \alpha^5} \right)
{\rm exp}(-\alpha),\label{subeq:c}
\end{equation}

\end{subequations}

\noindent\\
where ${\rm Ei}(-\alpha)$ is the exponential integral function.
The normalized coefficients (\ref{eq:4})-(\ref{eq:6}) for some
$\alpha$ values are shown in Table~\ref{tab:t1}.
Note that for $\alpha \gg 1$, eq.~(\ref{eq:4}) asymptotically
approaches unity, and for $\alpha \ll 1$,
approaches the expression of
${\tilde{\kappa}_e} = [5(2 \pi)^{1/2}/32] \alpha^{1/2}$ \cite{honda03}.
Another important point is that the heat capacity
tends to increase in high-temperature regimes,
up to twofold in the ultrarelativistic limit.
Making use of the implicit finite differential scheme,
we can numerically integrate the internal energy eqs.
(\ref{eq:2})-(\ref{eq:7}) as outlined in the Appendix.

In the LFFI context, of course, one can also construct
some kinds of kinetic simulation codes \cite{pukhov99}
to investigate asymmetric transport involving magnetic fields
\cite{pukhov97, honda00, sentoku02},
but their allowed spatiotemporal ranges are small.
Regarding the present scheme, the macroscopic transport properties
in {\it supersolid density} regions and the effects of
{\it a more realistic plasma gradient} as well, can be revealed
as a trade-off for missing the smaller spatiotemporal scales,
such as the Debye length and the plasma oscillation period.
Along the diffusion approximation adopted under geometrical constraint,
irradiating laser intensity $I_L$ is traced by
$v_g (\partial I_L/\partial r) = -\nu_{abs} I_L$,
where $v_g$ and $\nu_{abs}$ denote the group velocity of light
and the absorption coefficient, respectively \cite{kruer88}.
For simplicity, in the present simulations,
the laser light is set at normal incidence.
The ray deposits its own energy, when propagating through the corona,
and is then resonantly damped at the critical density.
Note that in eq.~(\ref{eq:1}) radiation pressure $P_{ph}$
is also taken into account.

Furthermore, we are concerned with the thermonuclear reactions of
deuterium-tritium (DT) and deuterium-deuterium (DD) fusion.
The rate equations for number density of the tritium $N_T$ and
deuterium $N_D$ \cite{huba94} were solved in a postprocessing manner.
Assuming the local self-heating due to charged particles,
one obtains the source terms of eqs.~(\ref{eq:2}) and (\ref{eq:3})
as functions of the heating power densities of
$S_{DT} \propto \left< v \sigma_{DT} \right> N_D N_T$
and $S_{DD} \propto \left< v \sigma_{DD} \right> N_D^2$ \cite{duderstadt82}.
Although these models are rather crude,
they still have merit for the investigation of
fundamental transport processes of electrons.
As shown later, it is instructive to compare the neutron yields
derived from various models of electron transport.

In the following, I demonstrate how a highly compressed fuel
can be heated due to electron transport, and ignited.
For convenience, 'case~A' and 'case~B' are used
as references for simulations with and without relativistic effects,
respectively.
The initial laser conditions and target parameters are shown in
Table~\ref{tab:t2}.
Hereafter, the target parameters are fixed.
The total mass of the target shell is on the order of that
presented in ref.~\cite{azechi91}, and
the initial aspect ratio of the shell is about
$\left(R/ \Delta R \right)_0 \sim 10$.
High compression of the shell,
maintaining a low entropy, is carried out by the
Gaussian pulse shaping of driver laser light.
The outer thin ablator of carbonized DT (CDT) is blown off
just before the deceleration phase, leaving the
dense compressed core of DT fuel.
Hydrodynamic instabilities concomitant with the low-entropy implosion
are omitted for the moment.

\begin{figure*}
\resizebox{150mm}{!}{\includegraphics{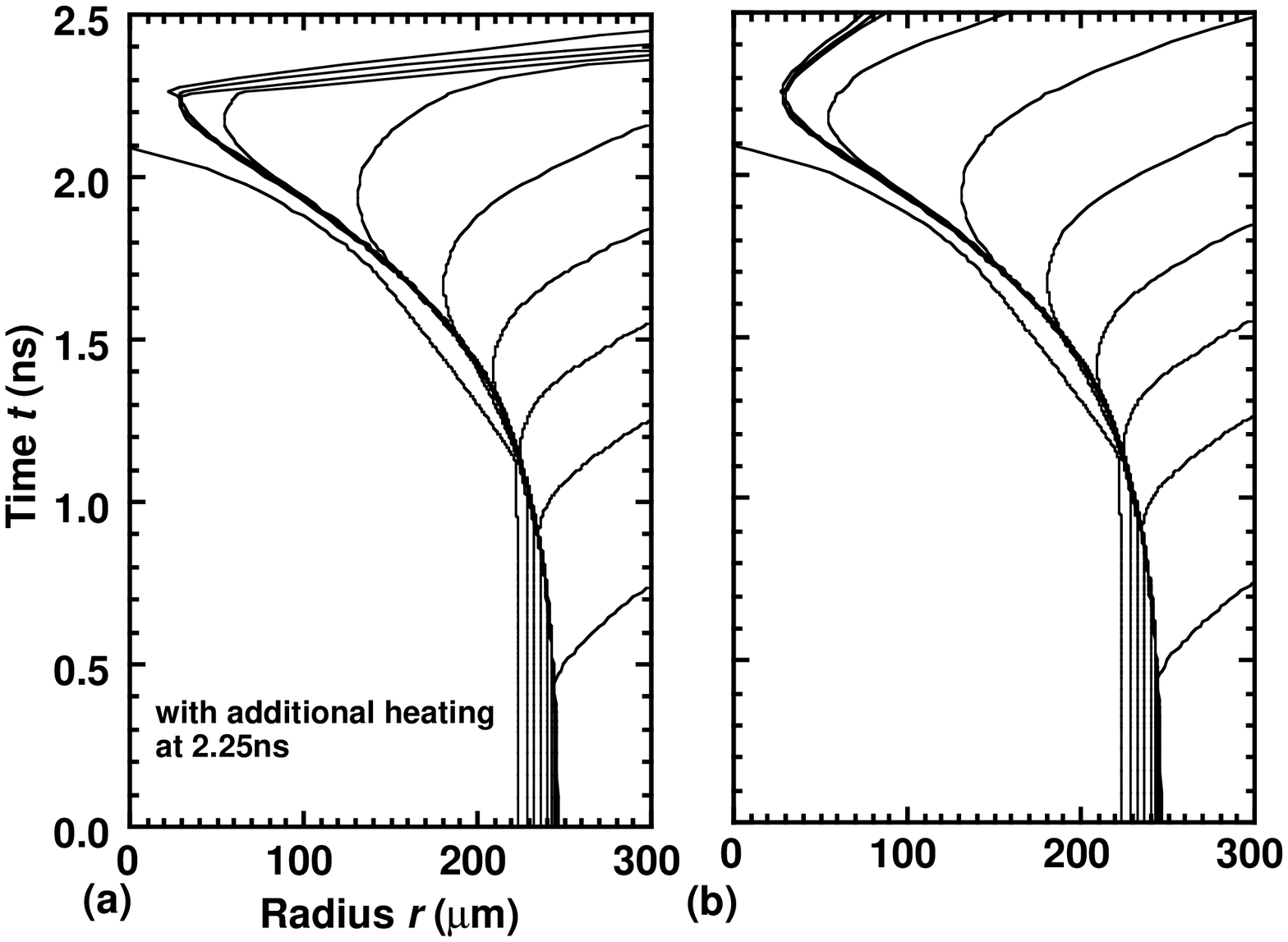}}
\caption{\label{fig:f1}
Flow diagrams of laser implosion
(a) with additional heating and
(b) without additional heating.
The figures have the same axes,
and the flow lines are displayed each tenth mesh.
At $t=2.250~{\rm ns}$ for (a), the additional heating power of
10~PW having the pulse duration of $\tau = 1~{\rm ps}$
is deposited at the relativistically corrected critical surface.
}
\end{figure*}

In Fig.~\ref{fig:f1}(a) for case~A, I show a flow diagram of
implosion with additional heating.
The total mesh number is $J=102$, and the flow lines are displayed
each tenth mesh.
At $t=2.25~{\rm ns}$, an additional $10$~PW ($10^{16}~{\rm W}$)
power with the pulse duration of $\tau = 1~{\rm ps}$ is
deposited at a relativistically modified cut-off density
which is defined by $n_{c\_rel}=n_c \sqrt{1 + I_L \lambda_L^2/
\left( 1.37 \times 10^{18}~{\rm W/cm^2-\mu m^2} \right) }$,
where $n_c = 9.97 \times 10^{20}~
\left( 1.06~{\rm \mu m}/\lambda_L \right)^2~{\rm cm}^{-3}$
is the nonrelativistic cut-off density.
In this case, the pulse intensity becomes about
$I_L \lambda_L^2 \sim 10^{18}~{\rm W/cm^2-\mu m^2}$.
One can see of the detonation shock propagating radially inwards,
as well as the explosion, associated with the off-center fast ignition.
The flow diagram without additional heating is shown in Fig.~\ref{fig:f1}(b).
It is confirmed that ignition does not occur in the standard implosion of the
small target when using a driver laser energy of about 4~kJ.

\begin{figure*}
\resizebox{150mm}{!}{\includegraphics{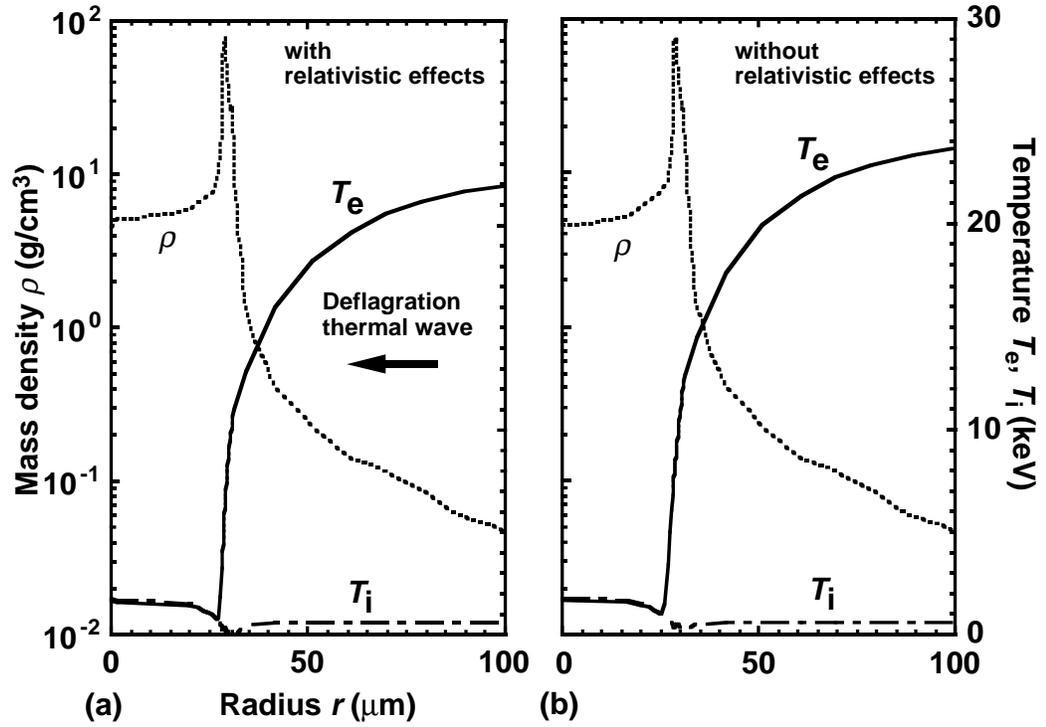}}
\caption{\label{fig:f2}
The spatial profiles of electron temperature (full curves),
ion temperature (dot-dashed curves), and mass density (dotted curves)
at $t=2.251~{\rm ns}$
(a) for case~A with relativistic effects and
(b) for case~B without relativistic effects.
The figures have the same axes.
}
\end{figure*}

The spatial profiles of plasma temperature and mass density at
$t = 2.251~{\rm ns}$, the moment that the additional heating power of
10~PW is switched off, are shown in Fig.~\ref{fig:f2} for cases~A and B.
One may notice that in the regions of $r = 0-35~{\rm \mu m}$,
the stagnating isobaric core where a central hot spot is surrounded
with cold dense plasma (DT main fuel) is well established
as a result of the low-entropy implosion.
The equilibrium static pressure reaches
$P = P_e + P_i > 10~{\rm Gbar}$.
It turns out that nonlinear propagation of the deflagration wave
directly heats the main fuel.
Here we define the average electron temperature of the main fuel as
$\left< T_e \right> = \int_{r_1}^{r_2} \rho T_e r^2 {\rm d}r/
\int_{r_1}^{r_2} \rho r^2 {\rm d}r$,
where $r_1$ and $r_2$ indicate the inner and outer radii where
$\rho = \rho_{\rm peak}/10$, respectively.
In case~B without relativistic corrections,
the average temperature is $\left< T_e \right> \simeq 9.9~{\rm keV}$.
On the other hand, in case~A, the heat flux inhibition
due to relativistic effects lowers the temperature to
$\left< T_e \right> \simeq 6.6~{\rm keV}$.
This reflects the factor
$\tilde \kappa_e/\tilde c_{ve} \simeq 0.75$
for $T_e \simeq 25~{\rm keV}$ in the tenuous corona
(see Table~\ref{tab:t1}).
The relativistic effects on energy transfer
between electrons and ions are small.

\begin{figure*}
\resizebox{150mm}{!}{\includegraphics{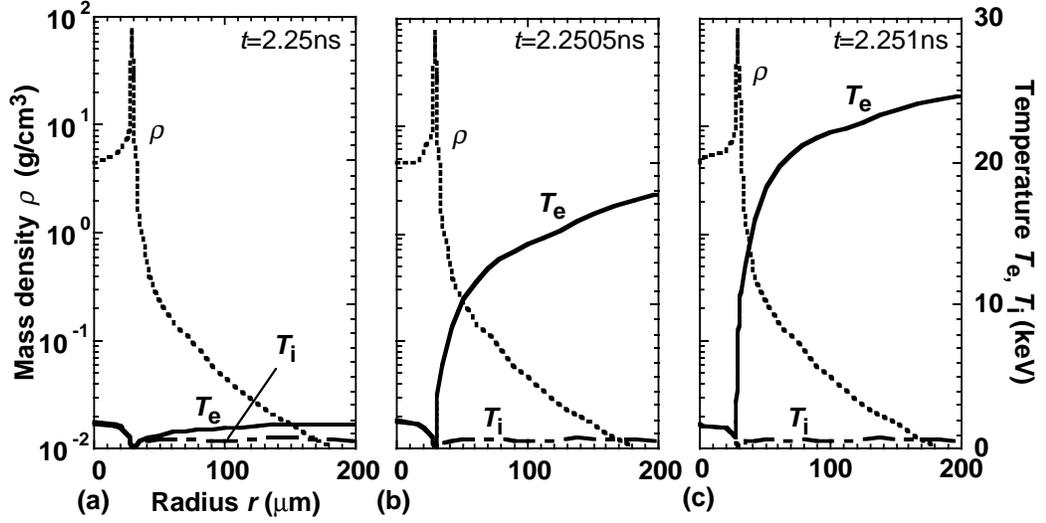}}
\caption{\label{fig:f3}
The spatial profiles of electron temperature (full curves),
ion temperature (dot-dashed curves), and mass density (dotted curves)
for case~A; (a) just before the pulse power injection,
(b) during the injection of 10~PW, and
(c) just after that. The figures have the same axes.
}
\end{figure*}

In Fig.~\ref{fig:f3} for case~A, I show the
spatial profiles of plasma temperature and mass density,
just before the irradiation of the ignitor pulse ($t = 2.25~{\rm ns}$),
during the irradiation ($t = 2.2505~{\rm ns}$),
and just after the irradiation ($t = 2.251~{\rm ns}$).
While the coronal electrons are rapidly heated up to
$T_e \sim 25~{\rm keV}$, the deflagration wavefront
slowly propagates in the cold dense region
where the heat capacity is very large,
and the steep temperature gradient at the wavefront
self-generates a longitudinal electric field
having a maximum value on the order of
magnitude of $E_{\rm max} \sim e^{-1}|\partial T_e/\partial r|
\sim 10^{10}~{\rm V/m}$ \cite{honda03}.
For the plasma parameters shown in Figs.~\ref{fig:f2} and \ref{fig:f3},
the heating rate of the DT fuel ions is estimated to be
${\rm d}T_i/{\rm d}t \sim 1-1.5~{\rm keV/ps}$,
when assuming the Coulomb logarithm of ${\rm ln\Lambda \sim 5}$.
The fuel can be, therefore, heated to the ignition temperature
until the stagnating core is disassembled.

\begin{figure*}
\resizebox{150mm}{!}{\includegraphics{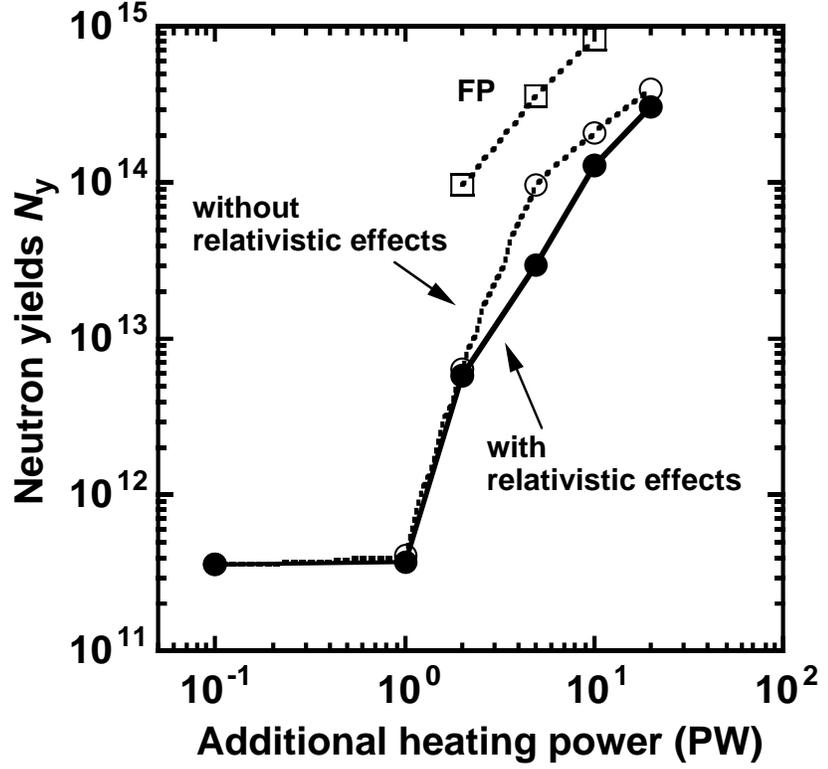}}
\caption{\label{fig:f4}
Additional heating power dependencies of neutron yields.
The full and open circles are for cases~A and B, respectively.
The target parameters and driver laser conditions for implosion
are the same as those given in Table~\ref{tab:t2}.
For comparison, the corresponding results of
the Fokker-Planck (FP) simulation are also shown (open squares).
For explanation see text.
}
\end{figure*}

Let us show the additional heating power dependencies of
neutron yields in Fig.~\ref{fig:f4} for both cases~A and B.
Within the range of $0.1-20~{\rm PW}$, an additional power
with the pulse duration of $\tau = 1~{\rm ps}$
is injected at $t = 2.250~{\rm ns}$.
It is found that in the present model the threshold power for
ignition seems to be about 1~PW, corresponding to the energy of 1~kJ.
The pulse power of about 20~PW leads to a considerable increase
in neutron yields, which are of the order of one thousand times that
in the case without additional heating.
The time-integrated bremsstrahlung loss is $60-70$~J at most.
It is noted that the flux inhibition degrades
the heating efficiency, thereby reducing the neutron yields,
particularly, when the additional power is, in this model,
in the range of $3-10$~PW.
The yields tend to be sensitive to the timing of the
irradiation of the ignitor pulse \cite{mima95, kodama02}.

The present predictions of the
neutron yields and the threshold power for ignition
may be rather pessimistic for determining the lower limit of the yields.
In order to argue this point, I show, in Fig.~\ref{fig:f4},
some results of the Fokker-Planck simulation
coupled with spherically imploding hydrodynamics \cite{mima95}.
In this calculated example,
the laser irradiation conditions and target parameters are similar to those
presented in Table~\ref{tab:t2}.
It is assumed that the absorbed energy of an ignitor pulse
is transfered by $50~\%$ into the high-energy tail electrons
of the temperature (variance) of $50~{\rm keV}$.
As shown in the figure,
the effects of nonlocal transport seem to increase the neutron yields,
since the hot tail electrons, which possess a longer mean-free path,
deeply penetrate into the compressed plasma, preheating the fuel.
It is noted that such nonlocality, as well as
anisotropy of energy deposition, could be an advantage to the LFFI.
The details should be clarified by
multidimensional fluid simulations including relativistic
kinetics in the future.

\appendix*

\section{Implicit difference of
relativistically corrected internal energy equations}

In this appendix, I briefly explain the numerical method to
integrate the relativistically corrected energy equations.
The implicit finite difference of eq.~(\ref{eq:2})
can be cast to

$$
{\frac {c_{ve,j+1/2}^* T_{e,j+1/2}^* - c_{ve,j+1/2}^n T_{e,j+1/2}^n}
{\Delta t^{n+1/2}}} =
-P_{e,j+1/2}^n {\frac {V_{j+1/2}^{n+1} - V_{j+1/2}^n}{\Delta t^{n+1/2}}}
$$
$$
-{\frac {4 \pi}{\Delta m_{j+1/2}}} [
\left( 1-\phi \right)
\left(
\kappa_{e,j+1}^* r_{j+1}^2
{\frac {T_{e,j+3/2}^* - T_{e,j+1/2}^*}{\Delta r_{j+1}^{n+1}}}
- \kappa_{e,j}^* r_j^2
{\frac {T_{e,j+1/2}^* - T_{e,j-1/2}^*}{\Delta r_j^{n+1}}}
\right)
$$
$$
+ \phi \left( \kappa_{e,j+1}^n r_{j+1}^2
{\frac {T_{e,j+3/2}^n - T_{e,j+1/2}^n}{\Delta r_{j+1}^n}}
- \kappa_{e,j}^n r_j^2
{\frac {T_{e,j+1/2}^n - T_{e,j-1/2}^n}{\Delta r_j^n}} \right) ]
$$
\begin{equation}
- \left( 1-\phi \right)
\omega_{ei,j+1/2}^* \left( T_{e,j+1/2}^* - T_{i,j+1/2}^n \right)
- \phi \omega_{ei,j+1/2}^n \left( T_{e,j+1/2}^n - T_{i,j+1/2}^n \right)
+ S_{j+1/2}^n,
\end{equation}

\noindent\\
where $r_j$~($j = 0,~1,~2,~...,J$) denotes the discrete positions of
fluid elements and $\phi$~($\leq 1/2$) is the implicit parameter,
e.g., the Crank-Nicholson scheme corresponds to the case of $\phi = 1/2$.
The last term on the right-hand side, $S_{j+1/2}^n$, represents generic
power sources and/or sinks.
Note that the coefficients $\kappa_e$, $\omega_{ei}$, and $c_{ve}$
depend upon the temperatures.

In the present simulation, the tridiagonal matrix (A.1) for $\phi=0$
is inverted by the cyclic reduction method \cite{press92}.
The temperature at the intermediate time steps
$T_{e,j+1/2}^*$ and the coefficients involving this are
iteratively advanced to the next ($T_{e,j+1/2}^{**}$).
When the condition of
$|T_{e,j+1/2}^{**} - T_{e,j+1/2}^*|/
\left( T_{e,j+1/2}^{**} + T_{e,j+1/2}^* \right) < \epsilon$
($\forall~j$) is satisfied for a small value of $\epsilon$,
the vector $T_{e,j+1/2}^{**}$ is replaced by $T_{e,j+1/2}^{n+1}$.
These procedures are repeated each time step,
whose increment $\Delta t^{n+1/2}$ is automatically changed,
invoking the Courant-Friedrich-Lewy condition.

\clearpage

\clearpage

\clearpage

\clearpage

\clearpage

\begin{table*}
\caption{\label{tab:t1}
The relativistic correction factors given by
eqs.~(\ref{eq:4})-(\ref{eq:7}) in the text.}
\begin{ruledtabular}
\begin{tabular}{clllll}
 $\alpha$~ &
 $T_e$~(${\rm MeV}$) &
 ~$\tilde \kappa_e$\footnotemark[1] &
 ~$\tilde \omega_{ei}$\footnotemark[2] &
 ~$\tilde c_{ve}$\footnotemark[3] &
 ~$\tilde \kappa_e/\tilde c_{ve}$\\ \hline
 $\ll 1$ & ~~$-$ & ~~$-$ & ~~$-$ & 2 & ~~$-$\\
 ~~~~0.05 & 10.2~~ & 0.087556 & 5.6084 & 1.9992 & 0.043796\\
 ~~0.1 & ~5.11 & 0.12373 & 3.9726 & 1.9968 & 0.061966\\
 ~~0.5 & ~1.02 & 0.27123 & 1.8510 & 1.9393 & 0.13986\\
 ~~1~~ & ~0.511 & 0.36792 & 1.4188 & 1.8343 & 0.20058\\
 ~~5~~ & ~0.102 & 0.62668 & 1.0528 & 1.3746 & 0.45589\\
 ~10~~ & ~0.0511 & 0.73598 & 1.0206 & 1.2156 & 0.60546\\
 ~20~~ & ~0.0256 & 0.83044 & 1.0084 & 1.1160 & 0.74420\\
 $\gg 1$ & ~~$-$ & 1 & 1 & 1 & 1\\
\end{tabular}
\end{ruledtabular}
\footnotetext[1]{Cited from ref.~\cite{honda98a} by Honda and Mima.}
\footnotetext[2]{Cited from ref.~\cite{beliaev56} by Beliaev and Budker.}
\footnotetext[3]{Cited from ref.~\cite{balescu75}
by Balescu and Paiva-Veretennicoff.}
\end{table*}

\clearpage

\begin{table*}
\caption{\label{tab:t2}
Simulation parameters of laser implosion including additional heating.}
\begin{ruledtabular}
\begin{tabular}{lllll}
 \multicolumn{3}{c}{Laser conditions}&\multicolumn{2}{c}{Target parameters}\\
 &Driver pulse&Ignitor pulse& \\ \hline
 ~Pulse shape: & Gaussian & Square
 & ~~~Inner radius: & 223.32~${\rm \mu m}$~(vacuum)\\
 ~Laser energy: & 4.1~kJ & 10~kJ\footnotemark[1]
 & ~~~${\rm D_{0.5}T_{0.5}}$: & 2.00~${\rm \mu m}$~(3 meshes)\\
 ~Wavelength: & 0.53~${\rm \mu m}$ & 1.06~${\rm \mu m}$
 & ~~~${\rm D_{0.5}T_{0.5}}$: & 16.92~${\rm \mu m}$~(42 meshes)\\
 ~Rise time: & 0.6~{\rm ns}\footnotemark[2] & ~~~$-$
 & ~~~${\rm C_{0.426}D_{0.534}T_{0.0153}}$:
 & 4.76~${\rm \mu m}$~(57 meshes)\\
 ~Pulse width: & 1.909~{\rm ns}\footnotemark[3]
 & 1.0~{\rm ps}\footnotemark[4] & & \\
 ~Fall time: & 0.5~{\rm ns}\footnotemark[5] & ~~~$-$ & & \\
 ~Peak power: & 2.5~TW & 10~PW\footnotemark[1] & & \\
\end{tabular}
\end{ruledtabular}
\footnotetext[1]{Variable between 0.1~kJ$-$20~kJ (100~TW$-$20~PW),
                 as seen in Fig.~\ref{fig:f4}.}
\footnotetext[2]{It follows the Gaussian pedestal of 1.2~{\rm ns}.}
\footnotetext[3]{FWHM, including the top flat part of 0.809~{\rm ns}.}
\footnotetext[4]{Switched on at $t=2.250$~{\rm ns}.}
\footnotetext[5]{Additional Gaussian pedestal of 1.0~{\rm ns}
                 follows its fall.}
\end{table*}


\begin{thebibliography}{99}

\bibitem{fisch87}
N.~J.~Fisch: Rev.~Mod.~Phys. {\bf 59} (1987) 175.

\bibitem{tabak94}
M.~Tabak, J.~Hammer, M.~E.~Glinsky, W.~L.~Kruer, S.~C.~Wilks, J.~Woodworth,
E~.M.~Campbell and M.~D.~Perry: Phys.~Plasmas {\bf 1} (1994) 1626.

\bibitem{pukhov97}
A.~Pukhov and J.~Meyer-ter-Vehn: Phys.~Rev.~Lett. {\bf 79} (1997) 2686.

\bibitem{mason98}
R.~J.~Mason and M.~Tabak: Phys.~Rev.~Lett. {\bf 80} (1998) 524.

\bibitem{caruso96}
A.~Caruso and V.~A.~Pais: Nucl. Fusion {\bf 36} (1996) 745.

\bibitem{atzeni97}
S.~Atzeni and M.~L.~Ciampi: Nucl.~Fusion {\bf 37} (1997) 1665.

\bibitem{piriz98a}
A.~R.~Piriz and M.~M.~Sanchez: Phys.~Plasmas {\bf 5} (1998) 2721.

\bibitem{piriz98b}
A.~R.~Piriz and M.~M.~Sanchez: Phys.~Plasmas {\bf 5} (1998) 4373.

\bibitem{honda00}
M.~Honda, J.~Meyer-ter-Vehn and A.~Pukhov:
Phys.~Rev.~Lett. {\bf 85} (2000) 2128.

\bibitem{sentoku02}
Y.~Sentoku, K.~Mima, Z.~M.~Sheng, P.~Kaw, K.~Nishihara and K.~Nishikawa:
Phys.~Rev.~E, {\bf 65} (2002) 046408.

\bibitem{tanaka00}
K.~A.~Tanaka, R.~Kodama and H.~Fujita: Phys.~Plasmas {\bf 7} (2000) 2014.

\bibitem{kodama01}
R.~Kodama, P.~A.~Norrey and K.~Mima: Nature {\bf 412} (2001) 798.

\bibitem{honda98a}
M.~Honda and K.~Mima: J.~Phys.~Soc.~Jpn {\bf 67} (1998) 3420.

\bibitem{honda98b}
M.~Honda and K.~Mima: Plasma~Phys.~Control.~Fus. {\bf 40} (1998) 1887.

\bibitem{longair94}
M.~S.~Longair: {\it High Energy Astrophysics}
(Cambridge Univ.,~Cambridge,~1994) 2nd ed., Vol.~2.

\bibitem{beliaev56}
S.~T.~Beliaev and G.~I.~Budker: Sov.~Phys.~Dokl. {\bf 1} (1956) 218.

\bibitem{balescu75}
R.~Balescu and I.~Paiva-Veretennicoff: Physica A {\bf 81} (1975) 17.

\bibitem{honda03}
M.~Honda: Phys.~Plasmas {\bf 10} (2003) 4177.

\bibitem{pukhov99}
A.~Pukhov: J.~Plasma~Phys. {\bf 61} (1999) 425.

\bibitem{kruer88}
W.~L.~Kruer: {\it The Physics of Laser Plasma Interactions}
(Wesley,~California,~1988).

\bibitem{huba94}
J.~D.~Huba: {\it NRL Plasma Formulary} (NRL, Washington~DC,~1994).

\bibitem{duderstadt82}
J.~J.~Duderstadt and G.~A.~Moses:
{\it Inertial Confinement Fusion} (Wiley,~New~York,~1982).

\bibitem{azechi91}
H.~Azechi, T.~Jitsuno and T.~Kanabe: Laser~Part.~Beams {\bf 9} (1991) 193.

\bibitem{mima95}
K.~Mima, M.~Honda, S.~Miyamoto and S.~Kato:
{\it Proc. 12th Int. Conf.
Laser Interaction and Related Plasma Phenomena,
Osaka, 1995} (AIP,~New~York,~1996), Vol.~369, p.179.

\bibitem{kodama02}
R.~Kodama and the Fast-Ignitor Consortium: Nature {\bf 418} (2002) 933.

\bibitem{press92}
W.~H.~Press, S.~A.~Teukolsky, W.~T.~Vetterling and B.~P.~Flannery:
{\it Numerical Recipes Vol.~1, Fortran Numerical Recipes}
(Cambridge Univ.,~Cambridge,~1992).

\end{thebibliography}
\end{document}